\documentclass[letterpaper]{JHEP3}

\usepackage{graphicx,epsf,amssymb,amsbsy,amsfonts,amssymb,amsmath,amsthm}

\newcommand{\bu}{\mathbf{u}}
\newcommand{\bw}{\mathbf{w}}
\newcommand{\btu}{\tilde{\mathbf{u}}}
\newcommand{\btw}{\tilde{\mathbf{w}}}

\newcommand{\e}{\epsilon}
\newcommand{\da}{{\dot{a}}}
\newcommand{\db}{{\dot{b}}}
\newcommand{\dc}{{\dot{c}}}

\def\sect#1{section~{\ref{#1}}}

\def\eqn#1{eq.~(\ref{#1})}
\def\eqns#1#2{eqs.~(\ref{#1}) and (\ref{#2})}
\def\fig#1{fig.~{\ref{#1}}}

\def\app#1{appendix~{\ref{#1}}}
\def\overalldelta#1{\delta^6\left(\sum_{i\in #1}p_i\right)\delta^4\left(\sum_{i\in #1}q_i\right)\delta^4\left(\sum_{i\in #1}\tilde{q}_i\right)}

\def\NeqFour{{{\cal N}=4}}

\def\nn{\nonumber}

\def\tree{{\rm tree}}

\def\E{\mathcal{E}}

\def\s#1{{      
        \setbox\charbox=\hbox{$#1$}
        \setbox\slabox=\hbox{$/$}
        \dimen\charbox=\ht\slabox
        \advance\dimen\charbox by -\dp\slabox
        \advance\dimen\charbox by -\ht\charbox
        \advance\dimen\charbox by \dp\charbox
        \divide\dimen\charbox by 2
        \raise-\dimen\charbox\hbox to \wd\charbox{\hss/\hss}
        \llap{$#1$} }}

\title{Dual Conformal Properties of Six-Dimensional Maximal Super Yang-Mills Amplitudes}
\author{Tristan Dennen\footnote{Email: tdennen@physics.ucla.edu}, and\;Yu-tin Huang\footnote{Email: yhuang@physics.ucla.edu}
\\ \\
\\
\it  Department of Physics and Astronomy,\\ UCLA,\\
Los Angeles, CA 90095-1547, USA\;}
\abstract{We demonstrate that the tree-level amplitudes of maximal super-Yang-Mills theory in six dimensions, when stripped of their overall momentum and supermomentum delta functions, are covariant with respect to the six-dimensional dual conformal group. Using the generalized unitarity method, we demonstrate that this property is also present for loop amplitudes. Since the six-dimensional amplitudes can be interpreted as massive four-dimensional ones, this implies that the six-dimensional symmetry is also present in the massively regulated four-dimensional $\mathcal{N}=4$ super-Yang-Mills amplitudes. }
\preprint{ UCLA-TEP-10-108}
\keywords{amplitudes, maximal super Yang-Mills, six dimensions, dual conformal}

\begin{document}

\section{Introduction}

Dual superconformal symmetry~\cite{DualConformal,DualConformal2} has played an important role in understanding the structure of planar four-dimensional $\mathcal{N}=4$ super-Yang-Mills (sYM) theory at both strong~\cite{Strong} and weak coupling~\cite{Weak,anomaly}. In particular, the closure of the original and dual superconformal symmetries forms an infinite-dimensional Yangian symmetry~\cite{Drummond:2009fd}, which has been extremely useful in determining the planar amplitudes of four-dimensional $\mathcal{N}=4$ sYM~\cite{Drummond:2008cr, Bern:2006ew, Korchemsky:2010ut, ArkaniHamed:2010kv}.  

Because the realization of this symmetry relies heavily on four-dimensional twistor variables~\cite{Hodges:2009hk, ArkaniHamed:2009vw, Mason:2009qx}, it is not immediately apparent how the symmetry behaves away from four dimensions. This is an important question because the loop amplitudes are infrared divergent and require regularization in four dimensions, and the dimensional regulator breaks the symmetry ~\cite{DualConformal2,anomaly,anomaly2}. Generically, any regularization scheme will result in either altering the dimensionality or the massless condition of the external momenta, both of which are essential to the definition of twistors. While one can modify the dual symmetry generators to account for massive regulators~\cite{Alday:2009zm}, thus making the symmetry exact, it is a priori not apparent that such a symmetry should exist without explicit calculation of the loop amplitudes, although it is expected to exist. 

To clarify these issues, six-dimensional four-point sYM multiloop amplitudes were recently set up~\cite{Bern:2010qa} using the six-dimensional spinor helicity formalism and on-shell superspace of refs.~\cite{CheungOConnell, DHS}. If one restricts the external momenta to a four-dimensional subspace, these should correspond to four-dimensional $\mathcal{N}=4$ sYM amplitudes with loop momenta continued to six dimensions. Interestingly, four-dimensional dual conformal symmetry can be used to restrict the form of the multiloop planar integrand, and at four points, this integrand can be straightforwardly extended to six dimensions. Furthermore, the four-dimensional dual conformal boost generator can be extended to incorporate a massive regulator~\cite{Drummond:2008cr}, which can be interpreted as extra-dimensional momenta.

In ref.~\cite{Bern:2010qa} it was conjectured that the six-dimensional maximal sYM $n$-point tree amplitude, when stripped of the momentum and supermomentum delta functions, transforms covariantly under dual conformal inversion. More precisely, the delta-function-independent part of the amplitude inverts with the same inversion weight on all external lines. The delta functions then introduce extra inversion weight due to the mismatch of mass dimensions of the momentum and supermomentum delta functions. This conjecture was checked explicitly against the simple four-point tree amplitude.  

In this paper, we will show that the conjecture holds for all $n\geq4$--point tree amplitudes. We will establish the proof by induction; assuming that the $(n-1)$--point amplitude inverts covariantly, via BCFW recursion relations~\cite{BCFW}, the $n$--point amplitude will invert in the same way. This proof follows a similar line given for the four-dimensional $\mathcal{N}=4$ sYM theory in ref.~\cite{Brandhuber:2008pf}. In addition, while this paper was in preparation, a tree-level proof of dual conformal symmetry of ten-dimensional sYM was given in ref.~\cite{CaronHuot:2010rj} 

At loop level, while it is expected that the six-dimensional loop integration measure spoils any dual conformal properties present at tree level, we can recover good behavior by restricting our attention to the integrand. Using the tree-level result, we will demonstrate that the multiloop planar integrands invert in the same fashion as in four dimensions; they are covariant with equal weight on all external lines, and with extra weight for the dual loop variables. We proceed by combining the tree-level result with the generalized unitarity method~\cite{UnitarityMethod} to show that all planar cuts, after restoring the cut propagators, invert uniformly, and thus the planar multiloop integrand inverts in the same way.

By restricting the loop integration to a four-dimensional subspace, the six-dimensional maximal sYM amplitudes can be interpreted as four-dimensional massively regulated $\NeqFour$ sYM amplitudes. Furthermore, the four-dimensional loop integration measure inverts with the precise weight to cancel the extra weight of dual loop variables in the integrand. Because ultraviolet divergences are absent in four dimensions, and the massive regulator does not break the six-dimensional dual conformal symmetry, one concludes that the regulated $\NeqFour$ amplitude will obey the exact symmetry. Assuming cut constructability of the loop amplitudes, which is expected for maximally supersymmetric Yang-Mills, this demonstrates that the dual conformal symmetry is an exact symmetry of the planar amplitude of massively regulated $\NeqFour$ theory.

This paper is organized as follows: In \sect{SixDimHelicitySubSection}, we give a brief review of the six-dimensional spinor helicity formalism, which provides a convenient set of on-shell variables for the representation of amplitudes. In \sect{OnShellInversion}, we introduce constraint equations which define dual coordinates in terms of the original on-shell coordinates. The dual conformal symmetry is then defined on these dual coordinates. Through the constraint equations, we are also able to define how the on-shell variables transform under dual conformal inversion. In \sect{TreeLevelProof}, we prove the covariance of the tree-level amplitudes via induction using BCFW recursion. In \sect{LoopProof}, we use the generalized unitarity method~\cite{UnitarityMethod} and the tree-level covariance to extend the result to loop level.

\section{Review of spinor helicity in six dimensions}
\label{SixDimHelicitySubSection}

The six-dimensional spinor helicity formalism laid out in refs.~\cite{CheungOConnell,DHS} provides a convenient set of variables to represent six-dimensional massless theories. For a discussion of the spinor helicity formalism in general dimensions see ref.~\cite{Boels:2009bv}. This formalism has been successfully applied to computations of loop amplitudes of the six-dimensional $\mathcal{N}=(1,1)$ sYM theory~\cite{Bern:2010qa,Brandhuber:2010mm}.\footnote{Besides (super) Yang-Mills amplitudes, these variables have also been used to analyze the $\mathcal{N}=(2,0)$ theory in ref.~\cite{Huang:2010rn}.} 

The on-shell degrees of freedom of each external particle are described by the variables
\begin{equation}
\left(\lambda_{i}^{Aa},\tilde{\lambda}_{iA\da},\eta_{ia},\tilde{\eta}_{i}^{\da}\right) \,,
\end{equation}
subject to the constraint
\begin{equation}
 \lambda_i^{Aa}\lambda_{ia}^B = \frac{1}{2}\epsilon^{ABCD}\tilde\lambda_{iC\da}\tilde\lambda_{iD}^{\da} \,.
\label{lambdaconstraint}
\end{equation}
The indices used here and throughout this paper represent various transformation properties, summarized in the following:
\begin{eqnarray}
\hbox{SU$^*$(4) Lorentz group labels:}&& \quad A,B,C, \cdots = 1,2,3,4 \nn\\
\hbox{SU(2)$\times$SU(2) little group labels:} && \quad a,b,c, \cdots =1,2 \,
  \hskip .3 cm \hbox{and} \hskip .3 cm \da, \db, \dc, \cdots = 1,2 \,.\nn\\
  \hbox{SO(5,1) vector labels:} && \quad \mu, \nu, \rho, \cdot\cdot\cdot = 0,1,2,\dots5\, \nn\\ 
\hbox{Particle/region labels:} && \quad i,j,k,r,s,l_i \,.
\end{eqnarray}
The bosonic variables $\left(\lambda_{i}^{Aa},\tilde{\lambda}_{iA\da}\right)$ are related to the momentum via 
\begin{align} 
& p_i^{AB}=\lambda_i^{A a}\,\epsilon_{ab}\lambda_i^{B b} \,, & 
& p_{iAB}=\tilde{\lambda}_{iA\da}\,\epsilon^{\da\db}\tilde{\lambda}_{iB\db} \,,
\label{VectorSpinor}
\end{align}
where the matrices $\e_{ab}$ and $\e^{\da\db}$ are the SU(2) little group metric, and the lowering and raising of the spinor variables are defined as 
\begin{align}
& \lambda_a = \e_{ab}\lambda^b\,,
& &\tilde\lambda^{\dot{a}} = \e^{\dot{a}\dot{b}}\tilde\lambda_{\dot{b}}\,,
\end{align}
with $\e_{12}=-1$, $\e^{12}=1$. One can see that \eqn{VectorSpinor} solves the massless condition 
\begin{equation}
p^2_i \propto \e_{ABCD}p^{AB}_ip^{CD}_i = 0.
\end{equation}
We represent the contraction between chiral and anti-chiral spinors as 
\begin{equation}
\lambda^{Aa}_{i}\tilde{\lambda}_{jA\db}=\langle i^a|j_{\db}].
\end{equation}

The fermionic variables $\eta_{ia},\tilde{\eta}_{i}^{\da}$ carry the information of the on-shell states of the maximal sYM theory. More explicitly, the on-shell states correspond to the coefficients of the $\eta_{ia},\tilde{\eta}_{i}^{\da}$ expansion of the scalar superfield,    
\begin{eqnarray}
\Phi(\eta,\tilde{\eta}) &=&
  \phi 
  + \chi^a \eta_a 
  + \phi'(\eta)^2 
  + \tilde{\chi}_{\da}\tilde{\eta}^{\da} 
  + g^a\,_{\da}\eta_a\tilde{\eta}^{\da}
  + \tilde{\psi}_{\da}(\eta)^2\tilde{\eta}^{\da} \nn\\
  && \null
  + \phi''(\tilde{\eta})^2
  + \psi^a\eta_a(\tilde{\eta})^2
  + \phi'''(\eta)^2(\tilde{\eta})^2 \,,
\end{eqnarray}
where $(\eta)^2 \equiv\tfrac{1}{2}\epsilon^{ab}\eta_b\eta_a$ and
$(\tilde{\eta})^2 \equiv\tfrac{1}{2}\epsilon_{\dot{a}\dot{b}}
\tilde{\eta}^{\db}\tilde{\eta}^{\da}$. Similar to the relationship between the spinor variables and the momenta $p_i$, one can solve the on-shell condition for supermomenta $q_i,\tilde{q}_i$ as
\begin{align}
&  q_i^A= \lambda^{Aa}_i\eta_{ia}\,,
&& \tilde{q}_{iA} = \tilde{\lambda}_{iA\da}\tilde{\eta}_{i}^{\da} \,.
\end{align}
For $n\ge 4$, the superamplitude can be written as a function of $(p_i,q_i,\tilde q_i)$, 
\begin{equation}
\mathcal{A}_n=
  \overalldelta{\E}
  f_n(p_i,q_i,\tilde{q}_i) \,,
\label{factorization}
\end{equation}
where here and throughout this paper, we use $\mathcal{E}$ to indicate the set of external legs, and the fermionic delta function is defined as
\begin{equation}
\delta^4 \left(\sum_{i\in\mathcal{E}} q_i^A\right) \equiv
\frac{1}{4!} \, \e_{BCDE} \, 
\left(\sum_{i\in\mathcal{E}} q_i^B\right) 
\left(\sum_{i\in\mathcal{E}} q_i^C\right) 
\left(\sum_{i\in\mathcal{E}} q_i^D\right) 
\left(\sum_{i\in\mathcal{E}}q_i^E\right) \,,
\end{equation}
and similarly for the antichiral $\tilde{q}_A$.

Due to the special kinematics of the three-point amplitude, one introduces additional SU(2) variables which are related to the usual spinor variables as~\cite{CheungOConnell}
\begin{align}
&\langle i^a| i+1_{\da}]=u_{i}^{a}\tilde{u}_{i+1\da}, && \langle i^a| i-1_{\da}]=-u_{i}^{a}\tilde{u}_{i-1\da} \,.
\label{udefine}
\end{align}
One also defines the pseudoinverse of $u$ as 
\begin{equation}
u_{ia}w_{ib}-u_{ib}w_{ia}=\epsilon_{ab} \,.
\label{wdefine}
\end{equation}

With these new variables, it can be shown that the three-point superamplitude is given by
\begin{eqnarray}
\null \hskip -1.3 cm 
 \mathcal{A}_{3}^{\tree}(1,2,3) &=&
-i \bigl(\bu_{1}\bu_{2}+
    \bu_{2}\bu_{3}+
    \bu_{3}\bu_{1}\bigr)
  \biggl(\sum_{i=1}^3 \bw_i\biggr)
  \bigl(\btu_{1}\btu_{2}+
    \btu_{2}\btu_{3}+
    \btu_{3}\btu_{1}\bigr)
    \biggl(\sum_{i=1}^3 \btw_i\biggr) \,,
\label{ThreePointSuperAmplitude}
\end{eqnarray}
where $\bu_i$ and $\bw_i$ are defined in terms of the $u_i^a$ and
$w_i^a$ as
\begin{eqnarray}
 \bu_i = u_i^a\eta_{ia}, \quad 
    \btu_i = \tilde{u}_{i\da}\tilde{\eta}_{i}^{\da}, \quad 
    \bw_i=w_i^a\eta_{ia}, \quad
    \btw_i = \tilde{w}_{i\da}\tilde{\eta}_{i}^{\da} \,.
\end{eqnarray}
%

\section{Dual conformal symmetry}
\label{OnShellInversion}

Dual conformal symmetry is a symmetry of the superamplitude that is made manifest by introducing dual (or region) variables subject to the following constraints~\cite{DualConformal2}:
\begin{eqnarray}
\nonumber 
&&(x_i-x_j)^{AB}=\lambda_{\{ij\}}^{Aa}\lambda^{B}_{\{ij\}a} \,, \hskip 1.0cm
  (x_i-x_j)_{AB}=\tilde{\lambda}_{\{ij\}A\dot{a}}\tilde{\lambda}_{\{ij\}B}^{\dot{a}} \,, \\
&&(\theta_i-\theta_j)^{A}=\lambda_{\{ij\}}^{Aa}\eta_{\{ij\}a} \,, \hskip 1.55cm
  (\tilde{\theta}_i-\tilde{\theta}_j)_{A}=\tilde{\lambda}_{\{ij\}A\dot{a}}\tilde{\eta}_{\{ij\}}^{\dot{a}},
\label{constraints}
\end{eqnarray}
where each leg is labeled by the indices $\{ij\}$ of the two adjacent regions, the order of which indicates the direction of momentum flow along the leg (for example, $p^\mu_{\{ij\}} = -p^\mu_{\{ji\}}$). For tree amplitudes, this notation is redundant since $j$ can always be chosen as $i+1$. However this prescription does not generalize to loop level, and thus we use a more general notation in anticipation of the multiloop discussion in \sect{LoopProof}. We will go back and forth between using indices $(i,j,\ldots)$ to label regions and to label legs; the meaning of the indices should be clear from the context. The superamplitude is viewed as a distribution on the full space $(x,\theta,\tilde{\theta},\lambda,\tilde{\lambda},\eta,\tilde{\eta})$, with delta function support on the constraint equations~(\ref{constraints}). The cyclic nature of the region variables then automatically enforces momentum and supermomentum conservation, and the first two equations in (\ref{constraints}) also imply \eqn{lambdaconstraint}.

To obtain the four-dimensional massive amplitudes, we break the six-dimensional spinors up into four-dimensional representations. Explicit details can be found in refs.~\cite{Bern:2010qa,Boels:2009bv}. Here we just note that the dual variables should also be broken into four-dimensional pieces and the fifth and sixth dimensional components. With $p_{\{ij\}}=(\check{p}_{\{ij\}},m_{\{ij\}},\tilde{m}_{\{ij\}})$, we have:
\begin{equation}
\check{x}_i-\check{x}_{j}=\check{p}_{\{ij\}},\;\;\;n_i-n_{j}=m_{\{ij\}},\;\;\;\tilde{n}_i-\tilde{n}_{j}=\tilde{m}_{\{ij\}}\,,
\end{equation}
where we use a check mark over a variable to indicate the components in the four-dimensional subspace. The physical mass squared is then $m_{\{ij\}}^2+\tilde{m}^2_{\{ij\}}$.

The dual conformal boost generator can be expressed as a composition of dual conformal inversions and translations,
\begin{equation}
  K^{\mu}=I\, P_{\mu}\, I\,,
\label{KPrelate}
\end{equation}
so we begin our discussion with the dual conformal inversion operator $I$. The inversion is defined on the Clifford algebra as
\begin{eqnarray}
  I[(\sigma^\mu)_{AB}] \equiv (\tilde{\sigma}_\mu)^{BA}\,, \hskip 1.0cm I[(\tilde{\sigma}^\mu)^{AB}] \equiv (\sigma_\mu)_{BA}\,,
\end{eqnarray}
and on the region variables as
\begin{equation}
  I[x_i^\mu] \equiv (x_i^{-1})_\mu = \frac{x_{i\mu}}{x_i^2}\,,\hskip 1.0cm 
  I[\theta_i^A] \equiv (x_i^{-1})_{AB} \theta_i^B \,, \hskip 1.0cm
  I[\tilde{\theta}_{iA}] \equiv (x_i^{-1})^{AB}\tilde{\theta}_{iB}\,.
\label{variableinversion}
\end{equation}

From the inversion of $x^\mu$, we also see that
\begin{eqnarray}
  I[(x_i-x_j)^{AB}] &=& (x_i^{-1})_{AC}(x_i-x_j)^{CD}(x_j^{-1})_{DB} \nn\\
  &=& (x_j^{-1})_{AC}(x_i-x_j)^{CD}(x_i^{-1})_{DB}\,,
\end{eqnarray}
and integration measures invert as
\begin{align}
  & I[d^6x_i] = (x_i^2)^{-6}d^6x_i \,,
  && I[d^4\theta_i] = (x_i^2)^2 d^4\theta_i \,,
  && I[d^4\tilde\theta_i] = (x_i^2)^2 d^4\tilde\theta_i\,.
\end{align}
With these definitions in hand, we can deduce the inversion properties of all of the other variables by requiring the invariance of the constraint equations~(\ref{constraints}) and the definitions of the $u$ and $w$ variables in \eqns{udefine}{wdefine}. We leave the proofs of these properties to \app{InversionAppendix} and collect the results here:
\begin{align}
  &I[\lambda_{\{ij\}a}^A] = 
      \frac{x_{iAB}\lambda_{\{ij\}}^{Ba}}{\sqrt{x_i^2x_j^2}} = 
      \frac{x_{jAB}\lambda_{\{ij\}}^{Ba}}{\sqrt{x_i^2x_j^2}} \,,
  &&I[\eta_{\{ij\}a}] = -\sqrt{\frac{x^2_i}{x^2_j}} 
      \Bigl(\eta^a_{\{ij\}}+(x^{-1}_i)_{AB} \, \theta^A_i\lambda_{\{ij\}}^{Ba}\Bigr) \,, \nn\\ 
  &I[\tilde{\lambda}_{\{ij\}A\da}] = 
      \frac{x_i^{AB}\tilde{\lambda}_{\{ij\}B}^{\da}}{\sqrt{x_i^2x_j^2}} =
      \frac{x_j^{AB}\tilde{\lambda}_{\{ij\}B}^{\da}}{\sqrt{x_i^2x_j^2}} \,,
  &&I[\tilde{\eta}_{\{ij\}}^{\da}] = -\sqrt{\frac{x^2_i}{x^2_j}}
      \Bigl( \tilde{\eta}_{\{ij\}\da} + (x_i^{-1})^{AB} \, \tilde{\theta}_{iA}\tilde{\lambda}_{\{ij\}B\da}\Bigr) \,, \nn\\ 
  &I[u_{ia}]=\frac{\beta u_i^a}{\sqrt{x^2_{i-1}}} \,, 
  &&I[w_{ia}]=-\frac{1}{\beta}\sqrt{x^2_{i-1}}w_i^{a} \,, \nn\\ 
  &I[\tilde{u}_{i\da}]=\frac{\tilde{u}_i^{\da}}{\beta\sqrt{x^2_{i-1}}} \,, 
  &&I[\tilde{w}_{i\da}]=-\beta\sqrt{x^2_{i-1}}\tilde{w}_i^{\da} \,,
\label{moreinversions}
\end{align}
where $\beta$ is an unfixed parameter that is irrelevant in our calculations.

Given these inversion rules, one can immediately deduce via \eqn{KPrelate} how each variable transforms under the dual conformal boost generator $K^\mu$. Alternatively, one can deduce the same information by requiring that the dual conformal boost generator respects all of the constraints in \eqn{constraints}. If we were to use the usual dual conformal boost generator in $x$ space,
\begin{eqnarray}
K^{\mu}=\sum_{i}\left(2\,x_i^{\mu}x_i^{\nu} - x_i^2\,\eta^{\mu\nu}\right)\frac{\partial}{\partial x_i^{\nu}}\,,
\end{eqnarray} 
the LHS of the definition of the $x_i$ in \eqn{constraints} would be nonzero under boosts, while the RHS would vanish. To correct this, we must add derivatives with respect to $\lambda$ and $\tilde\lambda$ to $K^\mu$. These new derivatives in turn would not be compatible with the definition of $\theta_i$, so we must also add $\theta$ and $\eta$ derivatives.
Requiring that all of the constraints in \eqn{constraints} are consistent with $K^\mu$ then yields 
\begin{eqnarray}
K^\mu&=&\sum_{i}\left[
\left(2 \, x^\mu_i x^\nu_i-x_i^2 \, \eta^{\mu\nu}\right)\frac{\partial}{\partial x_{i}^{\nu}}
+\theta^A_i(\sigma^{\mu})_{AB}x_i^{BC}\frac{\partial}{\partial \theta^C_i}
+\tilde{\theta}_{iA}(\tilde{\sigma}^{\mu})^{AB}x_{iBC}\frac{\partial}{\partial \tilde{\theta}_{iC}}\right] \nn\\
&+&\frac{1}{2}\sum_{\{jk\}} \left[\lambda^{Aa}_{\{jk\}}(\sigma^\mu)_{AB}(x_j+x_k)^{BC}\frac{\partial}{\partial \lambda^{Ca}_{\{jk\}}}
-(\theta_j+\theta_k)^A(\sigma^\mu)_{AB}\lambda_{\{jk\}a}^{B}\frac{\partial}{\partial \eta_{\{jk\}a}} \right.\nn\\
&&+\left.\tilde{\lambda}_{\{jk\}A\da}(\tilde{\sigma}^\mu)^{AB}(x_j+x_k)_{BC}\frac{\partial}{\partial \tilde{\lambda}_{\{jk\}C\da}}
-(\tilde{\theta}_j+\tilde{\theta}_k)_{A}(\tilde{\sigma}^{\mu})^{AB}\tilde{\lambda}^{\da}_{\{jk\}B}\frac{\partial}{\partial \tilde{\eta}_{\{jk\}}^{\da}}\right] \,,\nn\\
\end{eqnarray}
where $i$ runs over all regions, and $\{jk\}$ runs over all legs. The bosonic part of this generator was given in ref.~\cite{Bern:2010qa}. One can explicitly check that the infinitesimal transformations generated by this dual conformal boost generator match with those generated by \eqn{KPrelate}.

\section{Dual conformal properties of tree-level amplitudes}
\label{TreeLevelProof}

In this section, we show that the tree-level amplitudes of six-dimensional maximal sYM exhibit dual conformal covariance. In ref.~\cite{Bern:2010qa}, the four-point tree-level amplitude was shown to be covariant under dual conformal inversion,
\begin{equation}
I[\mathcal{A}_4^{\tree}]=(x^2_1)^2(x^2_1x^2_2x^2_3x^2_4)\mathcal{A}_4^{\tree}.
\end{equation}
Note that the extra factor $(x^2_1)^2$ relative to the four-dimensional result comes from the mismatch of the degrees of the momentum and supermomentum delta functions in six dimensions. In six dimensions, the momentum conservation delta function is of degree six instead of degree four as in four dimensions. Since the fermionic delta function is still of degree eight, there will be a mismatch in inversion weights of degree two in $(x^2_1)$. After separating out the delta functions from the rest of the amplitude,
\begin{equation}
\mathcal{A}_n^{\tree}=\overalldelta{\E} f_n\,,
\end{equation}
it was conjectured that the function $f_n$, for $n\ge4$, transforms as 
\begin{equation}
I[f_n]=\left(\prod_{i\in\E}x_i^2\right) f_n
\label{conjecture}
\end{equation}
under dual conformal inversion. We prove this by induction, utilizing the BCFW recursion relations~\cite{BCFW};  assuming that all $f_{m}$ transform as in \eqn{conjecture} for $4\le m<n$, each term in the BCFW recursive construction of $f_n$ will respect \eqn{conjecture}, and hence so will $f_n$. For the three-point amplitude, due to special kinematics, it is possible to consider the external momenta in a four-dimensional subspace. It is then conceivable that the four-dimensional dual conformal properties carry over to higher dimensions via covariance. However, closer inspection is warranted, because the polarization vectors of the gluons could point outside of the subspace. Furthermore, the six-dimensional three-point amplitude is not proportional to the supermomentum delta function, and hence $f_3$ cannot be defined.

Given that the function $f_n$ inverts as \eqn{conjecture}, acting with the dual conformal boost generator then gives 
\begin{equation}
K^{\mu}[f_n] =\left(\sum_{i\in\E} 2x^{\mu}_i\right)f_n.
\end{equation}

The above results can be rewritten for the massive amplitudes. In four-dimensional notation, the conformal inversion acts as 
\begin{align}
&I\left[\check{x}^\mu\right]=\frac{\check{x}_\mu}{x^2}\,,
&&I\left[n\right]=-\frac{n}{x^2}\,,
&&I\left[\tilde{n}\right]=-\frac{\tilde{n}}{x^2}\,,
\end{align}
where $x^2=\check{x}^2-n^2-\tilde{n}^2$. The massive amplitude then transforms under the dual conformal boost generators as
\begin{align}
& \check{K}^\mu [f_n] =\left(\sum_{i\in\E} 2\check{x}^{\hat{\mu}}_i\right)f_n \,,
&& K^{n} [f_n] =\left(\sum_{i\in\E} 2n_i\right)f_n \,,
&& K^{\tilde{n}} [f_n] =\left(\sum_{i\in\E} 2\tilde{n}_i\right)f_n \,.
\end{align}
The generator $\check{K}^\mu$ is closely related to the dual generator for the massively regulated amplitude~\cite{Alday:2009zm}. The bosonic dual variable part is 
\begin{equation}
\check{K}^\mu=\sum_{i}\left[2\,\check{x}_i^{\mu}\left(\check{x}_i^{\nu}\frac{\partial}{\partial \check{x}_i^{\nu}}
+n_i\frac{\partial}{\partial n_i}
+\tilde{n}_i\frac{\partial}{\partial \tilde{n}_i}\right) 
- x_i^2\frac{\partial}{\partial \check{x}_{i\mu}}\,\right]\,,
\end{equation}
while the bosonic part of the fifth and sixth components of $K^\mu$ is
\begin{eqnarray}
\nonumber K^n&=&\sum_{i}\left[2\,n_i\left(\check{x}_i^{\nu}\frac{\partial}{\partial \check{x}_i^{\nu}}
  +n_i\frac{\partial}{\partial n_i}
  +\tilde{n}_i\frac{\partial}{\partial \tilde{n}_i}\right) 
  + x_i^2\frac{\partial}{\partial n_{i}}\,\right] \,,\\
K^{\tilde{n}}&=&\sum_{i}\left[2\,\tilde{n}_i\left(\check{x}_i^{\nu}\frac{\partial}{\partial \check{x}_i^{\nu}}
  +n_i\frac{\partial}{\partial n_i}
  +\tilde{n}_i\frac{\partial}{\partial \tilde{n}_i}\right) 
  + x_i^2\frac{\partial}{\partial \tilde{n}_{i}}\,\right]\,.
\end{eqnarray}
Since the massive formulation is obtained straightforwardly from the six-dimensional formalism, from now on we will work with manifest six-dimensional covariance.

\subsection{The BCFW shift in dual coordinates.}

Taking the BCFW shift to be on legs $1$ and $n$, we have 
\begin{align}
p_{1}(z)&=p_1+zr \,,
  &q_{1}(z)&=q_1+zs \,,
  &\tilde{q}_{1}(z)&=\tilde{q}_1+z\tilde{s} \,,\nn\\
p_{n}(z)&=p_{n}-zr \,, 
  & q_{n}(z)&=q_{n}-zs \,, 
  & \tilde{q}_n(z)&=\tilde{q}_n-z\tilde{s} \,.
\end{align}
The precise forms of $r$, $s$ and $\tilde{s}$ are given in refs.~\cite{CheungOConnell, DHS}. For our purposes, it is sufficient to note that this implies a shift in only the dual coordinates $x_{1}$, $\theta_{1}$ and $\tilde{\theta}_{1}$, 
\begin{align}
p_{1}(z)&=x_{1}(z)-x_{2} \,,
  & q_{1}(z)&=\theta_{1}(z)-\theta_{2} \,,
  & \tilde{q}_1(z)&=\tilde{\theta}_1(z)-\tilde{\theta}_2 \,,\nn\\
p_{n}(z)&=x_{n}-x_{1}(z) \,, 
  & q_{n}(z)&=\theta_{n}-\theta_{1}(z) \,, 
  & \tilde{\theta}_n(z)&=\tilde{\theta}_n-\tilde{\theta}_1(z)\,,
\end{align}
where 
\begin{align}
x_{1}(z)&=x_{1}+zr\,, & \theta_{1}(z)&=\theta_{1}+zs\,, & \tilde{\theta}_1(z)&=\tilde{\theta}_1+z\tilde{s}\,.
\end{align}
Thus each BCFW term can be defined in a dual graph with just one shifted dual coordinate. We will denote the legs with shifted momentum by placing hats over the leg labels, while a hat over $x$ and $\theta$ is used for shifted regions. 

There are two types of BCFW diagrams, characterized by the presence or absence of a three-point subamplitude. We must consider each case separately, due to the fact that we cannot pull out an overall supermomentum conservation delta function from the three-point amplitude, and thus the three-point amplitude does not have the straightforward inversion of \eqn{conjecture}.

\subsection{BCFW diagrams without three-point subamplitudes}
\label{BCFWNoThree}

\begin{figure}
\begin{center}
  \includegraphics[scale=0.5]{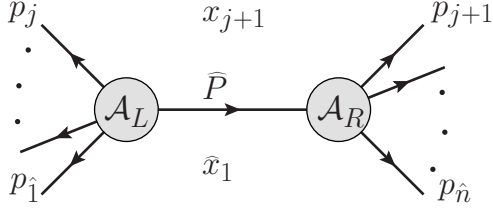}
  \caption{A BCFW diagram without three-point subamplitudes.}
  \label{BCFW}
\end{center}
\end{figure}

We first consider the case where there is no three-point subamplitude, as in \fig{BCFW}. 
The amplitudes on the left and right can be written as  
\begin{eqnarray}
\nonumber\mathcal{A}_L&=&\overalldelta{L}f_L(\hat{1},\cdots,j,\widehat{P}) \,,\\
\mathcal{A}_R&=&\overalldelta{R} f_R(-\widehat{P},j+1,\cdots,\hat{n}) \,.
\end{eqnarray} 
Each term in the BCFW recursion can then be written as 
\begin{equation}
\overalldelta{\E} f_{n}^{(j)}\,,
\end{equation}
where $f_{n}^{(j)}$ is the contribution to $f_n$ from the BCFW diagram labeled by $j$,
\begin{equation}
  f_{n}^{(j)} = \frac{i}{P^2}\int d^2\eta_Pd^2\tilde{\eta}_P 
  \,\delta^4\left(\sum_{i\in L} q_i\right)\delta^4\left(\sum_{i\in L} \tilde{q}_i\right) f_Lf_R\,.
\end{equation}

From the induction step, the functions $f_L$ and $f_R$ invert as 
\begin{eqnarray}
I\left[f_L\right]&=&\Bigl(\widehat{x}_1^2 x_2^2 \cdots x_{j+1}^2\Bigr)f_L\,, \nn\\
I\left[f_R\right]&=&\Bigl(x_{j+1}^2 \cdots x_n^2 \widehat{x}_1^2\Bigr)f_R\,.
\label{finversions}
\end{eqnarray} 
The propagator in $f_{n}^{(j)}$ has a simple inversion, given by
\begin{equation}
I\left[\frac{1}{P^2}\right]=I\left[\frac{1}{x^2_{1,j+1}}\right]=\frac{x^2_{1}x^2_{j+1}}{x^2_{1,j+1}}\,,
\label{propagatorinversion}
\end{equation}
so the only remaining piece of $f_{n}^{(j)}$ is the fermionic integral. Since the fermionic delta function is of degree eight, the fermionic integral can be completely localized by the delta functions, and the $\eta_P,\tilde{\eta}_P$s in $f_L,f_R$ will be replaced by the solution of the delta functions. The replacement does not affect the inversion properties of $f_L,f_R$ because it simply amounts to the use of supermomentum conservation. The integral has been shown previously~\cite{Bern:2010qa} to give
\begin{eqnarray}
\nonumber \int d^2\eta_Pd^2\tilde{\eta}_P \,
\delta^4\left(\sum_{i\in L} q_i\right)\delta^4\left(\sum_{i\in L} \tilde{q}_i\right)
 &=&\left(\widehat{\theta}_1-\theta_{j+1}\right)^A \tilde{\lambda}_{\widehat{P}A\da}\tilde{\lambda}_{\widehat{P}B}^{\da} \left(\widehat{\theta}_1-\theta_{j+1}\right)^B \\
&\phantom{=}&\times\biggl(\widehat{\tilde{\theta}}_1-\tilde{\theta}_{j+1}\biggr)_C \lambda_{\widehat{P}}^{Ca}\lambda_{\widehat{P}a}^D \biggl(\widehat{\tilde{\theta}}_1-\tilde{\theta}_{j+1}\biggr)_D\,.
\end{eqnarray}
Note that we do not write $f_L$ and $f_R$ in the integral because they are independent of $\eta_P,\tilde{\eta}_P$ after the replacement.
To see how this expression inverts, we use \eqns{variableinversion}{moreinversions} on each factor, such as 
\begin{eqnarray}
\nonumber I\left[\left(\widehat\theta_{1}-\theta_{j+1}\right)^A\tilde\lambda_{\widehat{P}A\da}\right]&=&
  -\frac{1}{\sqrt{\widehat{x}^2_{1}x^2_{j+1}}} \left(\widehat\theta_1^B(\widehat{x}^{-1}_1)_{BA}\widehat{x}_1^{AC}\tilde\lambda^{\da}_{\widehat{P}C} - \theta_{j+1}^B(x^{-1}_{j+1})_{BA}x^{AC}_{j+1}\tilde\lambda^{\da}_{\widehat{P}C}\right)\\
&=&-\frac{1}{\sqrt{\widehat{x}^2_1 x^2_{j+1}}} \left(\widehat\theta_1-\theta_{j+1}\right)^A\tilde\lambda^{\da}_{\widehat{P}A}\,.
\end{eqnarray}
Doing the same for the other factors, we find
\begin{eqnarray}
&& I\left[\int d^2\eta_Pd^2\tilde{\eta}_P \, \delta^4\left(\sum_{i\in L} q_i\right)\delta^4\left(\sum_{i\in L} \tilde{q}_i\right)\right]  \nn\\
&&\hskip 2.0cm =\frac{1}{(\widehat{x}^2_1 x^2_{j+1})^2}\int d^2\eta_Pd^2\tilde{\eta}_P \,  \delta^4\left(\sum_{i\in L} q_i\right)\delta^4\left(\sum_{i\in L} \tilde{q}_i\right)
\label{integralinversionA}
\end{eqnarray}

Combining equations~(\ref{finversions}),~(\ref{propagatorinversion}) and (\ref{integralinversionA}), we arrive at the desired result 
\begin{eqnarray}
I\left[f_{n}^{(j)}\right]=\left(\prod_{i\in \mathcal{E}}x^2_{i}\right)f_{n}^{(j)} \,.
\end{eqnarray}
%

\subsection{BCFW diagrams with a three-point subamplitude}

\begin{figure}
\begin{center}
  \includegraphics[scale=0.5]{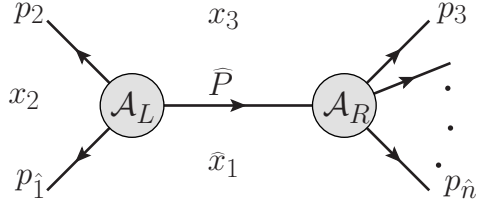}
  \caption{A BCFW diagram with a three-point subamplitude.}
  \label{BCFW2}
\end{center}
\end{figure}

To make a statement about the inversion weight of the entire $n$-point amplitude, we must also consider the BCFW terms which contain a three-point subamplitude, as shown in \fig{BCFW2}. 
It was shown in ref.~\cite{Bern:2010qa} that the contribution of such a diagram is given as,  
\begin{eqnarray}
&&\int d^2\eta_P d^2\tilde{\eta}_P \, \mathcal{A}_3\frac{i}{P^2}\mathcal{A}_{n-1}\\
\nonumber&=&-\overalldelta{\E}
\left(\bu_{2}-\bu_{\hat1}\right)\left(\btu_2-\btu_{\hat1}\right)\frac{1}{P^2}f_{n-1}\,,
\end{eqnarray}
where $f_{n-1}$ has been rewritten completely in terms of external leg variables by using the substitutions $q_{\widehat{P}}=-q_{\hat{1}}-q_{2}$ etc.  
Hence,
\begin{equation}
f_{n}^{(2)}=-\left(\bu_{2}-\bu_{\hat1}\right)\left(\btu_{2}-\btu_{\hat1}\right) \frac{1}{P^2} f_{n-1}\,.
\end{equation}
The inversion of $P^2$ and $f_{n-1}$ here are straightforward, and we are left with the remaining factors involving $\bu$ and $\btu$.
We consider the inversion of $\left(\bu_{2}-\bu_{\hat1}\right)$ in detail. After applying \eqn{moreinversions}, we get
\begin{eqnarray}
I\left[\bu_{2}-\bu_{\hat1}\right]&=&
  -\sqrt{\frac{x^2_{2}}{\widehat{x}^2_{1}x^2_{3}}} \, \beta u_{2a}\left(\eta^a_{2}+(x^{-1}_{2})_{AB}\theta^A_{2}\lambda_{2}^{Ba}\right) \nn\\
&&+\sqrt{\frac{\widehat{x}^2_1}{x^2_2 x^2_3}} \, \beta u_{\hat1a}\left(\eta^a_{\hat1}+(\widehat{x}^{-1}_{1})_{AB}\widehat\theta^A_{1}\lambda_{\hat{1}}^{Ba}\right)\,.
\label{invertuterm}
\end{eqnarray}
We can combine the $\theta$-dependent terms in the above equation as
\begin{eqnarray}
&&\frac{\beta}{\sqrt{\widehat{x}^2_1 x^2_2 x^2_3}} 
  \left(-u_{2a}\,x_{2AB}\,\theta^A_2\lambda_2^{Ba}
  +u_{\hat1a}\,\widehat{x}_{1AB}\,\widehat\theta^A_1\lambda_{\hat1}^{Ba}\right) \nn\\
&&\hskip2.0cm =\frac{-\beta}{2\sqrt{\widehat{x}^2_1 x^2_2 x^2_3}}
  \, u_{\hat1a}(\widehat{x}_1+x_2)_{AB}(\theta^A_{2}-\widehat\theta^A_{1})\lambda_{\hat1}^{Ba} \nn\\
&&\hskip2.0cm =\frac{\beta}{2\sqrt{\widehat{x}^2_1 x^2_2 x^2_3}}
  \, u_{\hat1a}(\widehat{x}_1+x_2)_{AB}\lambda^{Ab}_{\hat1}\lambda_{\hat1}^{Ba}\eta_{\hat1b} \nn\\
&&\hskip2.0cm =\frac{-\beta}{4\sqrt{\widehat{x}^2_1 x^2_2 x^2_3}}
  \, (\widehat{x}_1-x_2)^{AB} \, (\widehat{x}_1+x_2)_{AB} \, \bu_{\hat1} \nn\\
&&\hskip2.0cm =\beta \, \bu_{\hat1} 
  \left(\sqrt{\frac{\widehat{x}^2_1}{x^2_2 x^2_3}} - \sqrt{\frac{x^2_2}{\widehat{x}^2_1 x^2_3}}\right) \,,
\end{eqnarray}
where in the second line we have used $\widehat{x}_{1AB}\,\lambda_{\hat1}^{Ba} = x_{2AB}\,\lambda_{\hat1}^{Ba}$ and $u_{\hat1a}\lambda_{\hat1}^{Ba} = u_{2a}\lambda_{2}^{Ba}$. 
Putting this back into \eqn{invertuterm}, we arrive at 
\begin{equation}
I\left[(\bu_{2}-\bu_{\hat1})\right]= \beta
  \sqrt{\frac{x^2_2}{\widehat{x}^2_1 x^2_3}} (\bu_2-\bu_{\hat1}) \,.
\end{equation}
The inversion of the antichiral factor $(\btu_2-\btu_{\hat1})$ behaves in the same way, except that $\beta$ appears in the denominator.
Thus, putting everything together, we have 
\begin{equation}
I\left[f_{n}^{(2)}\right]=\biggl(\frac{x^2_2}{\widehat{x}^2_1 x^2_3}\biggr) \left(x^2_1 x^2_3\right)
  \left(\widehat{x}^2_1 x^2_3 \cdots x^2_n \right) f_n^{(2)}
  =\left(\prod_{i\in\E}x^2_{i}\right)f_{n}^{(2)}.
\label{f2invert}
\end{equation}
This completes the proof of \eqn{conjecture}. In the next section, we turn our attention to planar multiloop amplitudes.

\section{Loop amplitudes through unitarity cuts}
\label{LoopProof}

In this section, we will demonstrate to all loop orders that the $L$--loop planar integrand is covariant under inversion in the following way:
\begin{equation}
I\left[\mathcal{I}_n^{L}\right]= \left(\prod_{i\in\E} x_i^2\right)     \left(\prod_{i=1}^L (x_{l_i}^2)^4\right) \mathcal{I}_n^{L} \,,
\label{loopinvert}
\end{equation}
where the integrand is defined with respect to the amplitude as
\begin{eqnarray}
  \mathcal{A}_n^{L} =  \overalldelta{\E}
    \int \Biggl( \prod_{i=1}^{L}d^6 x_{l_i} \Biggr) \mathcal{I}_n^{L}\,.
\label{integranddefinition}
\end{eqnarray}
Because we are focusing on the integrand itself, there are extra loop region weights $(x_{l_i}^2)^4$. This is the same result as in four dimensions, although in six dimensions the loop integration measure inverts with weight $(x_{l_i}^2)^{-6}$, which does not exactly cancel the weight of the integrand. Therefore, the amplitude after integration will not be covariant unless the integral is restricted to four dimensions, which, as we have discussed, is the case when interpreting the extra two dimensions as a massive regulator~\cite{Alday:2009zm}. 

Our approach to \eqn{loopinvert} is to study the inversion properties of unitarity cuts of the amplitude. In the unitarity method, we are required to perform state sums across the cut propagators, which is achieved by integrating the Grassmann variables $\eta_{l_i},\tilde{\eta}_{l_i}$ of the cut lines. Since the tree amplitudes contributing to the cuts have definite inversion properties, we only need to understand how the $\eta_{l_i},\tilde{\eta}_{l_i}$ integration modifies the inversion weight. 

\begin{figure}
\begin{center}
  \includegraphics[scale=0.5]{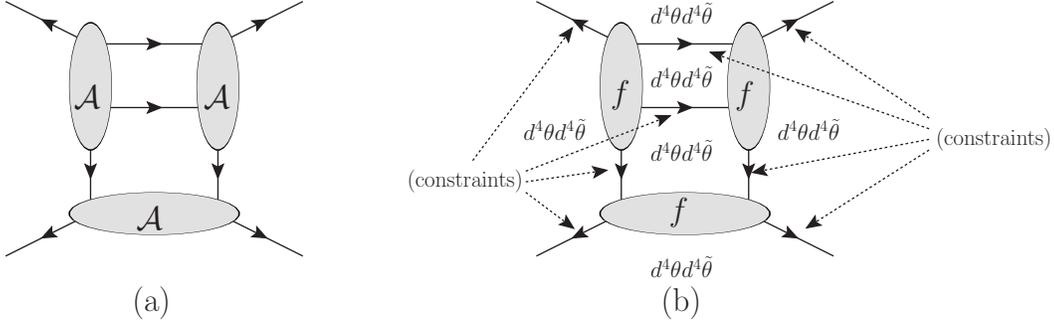}
  \caption{A cut of the two-loop four-point amplitude. (a) In the usual expression of the cut, this diagram is dressed with a tree-level amplitude for each blob and a state sum over each internal line. (b) As discussed in the text, for planar cuts this is equivalent to dressing the diagram with an $f$ function for each blob, introducing the dual variable constraints for every line, and integrating over the dual $\theta$ variables of every region. Finally, a state sum over each internal line is performed. One can check that the dressing of (b) contains $8\times 8=64$ fermionic delta functions and $5\times8=40$ integrations over $\theta$ (because one of the six regions is fixed by the shift symmetry), leaving $24$ unintegrated fermionic delta functions, which are exactly the supermomentum conservation of the subamplitudes in dressing (a).}
  \label{loopexample}
\end{center}
\end{figure}

To make statements about inversion properties, it is more natural to express everything in terms of dual variables than in terms of $\eta$ and $\lambda$. We therefore trade the supersum $\eta$ integrals for $\theta$ integrals. Suppose a cut not containing any three-point subamplitudes has an internal line between regions $i$ and $j$. The supersum across this line is expressed as an integral with measure $d^2\eta_{\{ij\}} d^2\tilde{\eta}_{\{ij\}}$. The transformation to dual coordinates is achieved by inserting 1 into the cut in a particular way, given by
\begin{eqnarray}
  \mathcal{A}_n^{L}\Bigr|_{\hbox{\footnotesize{cut}}} &=& \int 
    \prod_{\{ij\}} d^2\eta_{\{ij\}}d^2\tilde\eta_{\{ij\}} \times
    \mathcal{A}^{\tree}_{(1)}\mathcal{A}^{\tree}_{(2)}\mathcal{A}^{\tree}_{(3)}\ldots\mathcal{A}^{\tree}_{(m)} \nn\\
  &=& \int 
    \prod_{\{ij\}} d^2\eta_{\{ij\}}d^2\tilde\eta_{\{ij\}} \times
    \prod_{\alpha} \delta^4\left(\sum_{k \in\alpha} q_k\right) \delta^4 \left(\sum_{k\in\alpha} \tilde q_k\right) f_{\alpha}  \nn\\
  &=& \int
    \prod_{\{ij\}} d^2\eta_{\{ij\}}d^2\tilde\eta_{\{ij\}} \times
    \prod_k d^4\theta_kd^4\tilde\theta_k \times 
    \prod_\alpha f_\alpha \nn\\
  &&\times\prod_{\{rs\}}\delta^4\left(\theta_r^A-\theta_s^A-\lambda_{\{rs\}}^{Aa}\eta_{\{rs\}a}\right)
    \delta^4\left(\tilde\theta_{rB}-\tilde\theta_{sB}-\tilde\lambda_{\{rs\}B\da} \tilde\eta_{\{rs\}}^{\da}\right) \,,
\label{loopamplitude}
\end{eqnarray}
where the product over $\{ij\}$ runs over all internal cut lines, the product over $k$ runs over all regions, the product over $\{rs\}$ runs over all lines, and the product over $\alpha$ runs over the tree subamplitudes. The first two lines of this equality are the definition of the cut, where we have ignored the momentum conservation delta functions on the subamplitudes, because they combine straightforwardly into an overall momentum conservation when cut conditions are relaxed and loop integrals are replaced. Because the integrand in the third and fourth lines has a shift symmetry in the $\theta$ variables, the measure $\prod d^4\theta$ is understood to include only $(F-1)$ of the regions, where $F=n+L$ is the total number of regions in the graph. An explicit example for the two-loop four-point amplitude is given schematically in \fig{loopexample}. It does not matter how we fix the symmetry in the measure; our choice will only affect the overall supermomentum delta function, which does not contribute to the conjectured transformation \eqn{loopinvert}. We therefore leave this detail implicit.

To see the equality of \eqn{loopamplitude}, note that we can pull the subamplitude supermomentum delta functions out of the $\theta$ delta functions in the fourth line, leaving behind $(P-V)$ delta functions to be used for localizing the $\theta$ integrals, where $P$ is the number of lines in the graph, and $V$ is the number of subamplitudes. Because there are $(F-1)$ of the $\theta$ integrals, the leftover delta functions saturate the integral when $F-1=P-V$, which is indeed the case for planar graphs.

We can now use the $\theta$ delta functions to eliminate all explicit $\eta$ dependence from each $f_\alpha$, so that the entire $\eta$ dependence of the cut appears in the form
\begin{eqnarray}
  \int d^2\eta_{\{ij\}} \delta^4\left(\theta_i^A-\theta_j^A-\lambda_{\{ij\}}^{Aa}\eta_{\{ij\}a}\right) \,.
\end{eqnarray}
This performs the chiral half of the supersum across the line between regions $i$ and $j$. The antichiral half of the supersum is completely analogous, so we leave it out. The integration over $\eta_{\{ij\}}$ thus contributes
\begin{eqnarray}
  \theta_{ij}\cdot x_{ij}\cdot \theta_{ij} \equiv (\theta_i-\theta_j)^A (x_i-x_j)_{AB} (\theta_i-\theta_j)^B \,.
\end{eqnarray}
We demonstrated in \sect{BCFWNoThree} that this factor inverts with weight $(x_i^2x_j^2)^{-1}$.

Returning to the cut in \eqn{loopamplitude}, the result of doing the $\eta$ integrals is
\begin{eqnarray}
  \mathcal{A}_n^{L}\Bigr|_{\hbox{\footnotesize{cut}}} &=& \int
    \prod_k d^4\theta_kd^4\tilde\theta_k \times 
    \prod_\alpha f_\alpha^0 \times
    \prod_{\{ij\}} (\theta_{ij}\cdot x_{ij} \cdot \theta_{ij}) 
      (\tilde\theta_{ij} \cdot x_{ij} \cdot \tilde\theta_{ij})\nn\\
  &&\times\prod_{\{rs\}}\delta^4\left(\theta_r^A-\theta_s^A-\lambda_{\{rs\}}^{Aa}\eta_{\{rs\}a}\right)
    \delta^4\left(\tilde\theta_{rA}-\tilde\theta_{sA}-\tilde\lambda_{\{rs\}A\da} \tilde\eta_{\{rs\}}^{\da}\right) \,,
\end{eqnarray}
where now $\{rs\}$ only runs over the external lines. An overall supermomentum delta function pulls out, leaving $(n-1)$ delta functions of each chirality, which completely saturate the $\theta$ integrations over the external regions (this also takes care of the shift symmetry detail). We are finally left with
\begin{eqnarray}
  \mathcal{A}_n^{L}\Bigr|_{\hbox{\footnotesize{cut}}} &=& \overalldelta{\E} \nn\\
  &&\times   \int \left(\prod_k d^4 \theta_k d^4 \tilde\theta_k \right) 
  \left( \prod_{\{ij\}} (\theta_{ij}\cdot x_{ij}\cdot \theta_{ij})  (\tilde\theta_{ij}\cdot x_{ij} \cdot \tilde\theta_{ij})\right)
  \prod_\alpha f_\alpha \,, 
\label{cutintegrand}
\end{eqnarray}
where the product over $k$ now runs only over the internal regions, and we have replaced the overall momentum conservation.

We are now in a position to formulate a set of diagrammatic rules for inverting the cut, after restoring the cut propagators. Because each piece of the second line of \eqn{cutintegrand} inverts covariantly, the cut inverts to itself multiplied by an overall prefactor (not considering the inversion of the overall delta functions). To calculate the prefactor for a given cut, we have the following rules:
\begin{itemize}
  \item For every loop region $k$, the $\theta_k,\tilde\theta_k$ measure contributes a factor $(x_k^2)^4$.
  \item Each internal leg $\{ij\}$ contributes $(x_i^2x_j^2)^{-1}$, where a factor of $x_i^2 x_j^2$ comes from the cut propagator, and a factor of $(x_i^2x_j^2)^{-2}$ comes from $(\theta_{ij}\cdot x_{ij}\cdot\theta_{ij})(\tilde\theta_{ij}\cdot x_{ij}\cdot\tilde\theta_{ij})$
  \item Each tree-level subamplitude contributes $\prod_i x_i^2$, where $i$ runs over all regions adjacent to the tree.
\end{itemize}
Given a region $i$, it is straightforward to invert these rules to figure out what power of $x_i^2$ appears in the prefactor. If $i$ is an external region, $x_i^2$ must appear to the power $(\rho_i-\sigma_i)$, where $\rho_i$ and $\sigma_i$ are the number of tree-level subamplitudes and the number of internal propagators, respectively, adjacent to region $i$. Each external region necessarily borders one fewer of the internal propagators than the subamplitudes, so the external regions each give $x_i^2$. If, on the other hand, $i$ is an internal region, then $x_i^2$ appears to the power $(\rho_i-\sigma_i+4)$. All internal regions necessarily border the same number of internal propagators as subamplitudes, so the internal regions each give $(x_i^2)^4$. Therefore, we have reached the result that each planar cut with no three-point subamplitudes inverts with the prefactor
\begin{equation}
  \left(\prod_{i\in\E} x_i^2\right)\left(\prod_{i=1}^{L} (x_{l_i}^2)^4\right)\,,
\end{equation}
after the cut propagators have been restored, and not including the overall momentum and supermomentum conservation. It is not difficult to extend this result to cuts involving three-point subamplitudes. The supersum between a three-point subamplitude and another subamplitude in a cut proceeds in the same way as sewing a three-point tree in BCFW. The resulting merged subamplitudes then invert as in \eqn{f2invert}.

Because all cuts invert in exactly the same way, and the correct amplitude must satisfy all generalized unitarity cuts, we conclude that the $L$--loop integrand inverts as
\begin{equation}
  I\left[\mathcal{I}_n^L\right] = \left(\prod_{i\in\E} x_i^2\right)\left(\prod_{i=1}^{L} (x_{l_i}^2)^4\right) \mathcal{I}_n^L\,.
\label{conclusion}
\end{equation}
For a recent discussion of the transition from cuts to the amplitude, see ref.~\cite{Bern:2010tq}. Note that bubbles on external lines are not cut detectable, so they potentially violate \eqn{conclusion}. However, because this is the maximally supersymmetric theory, we do not expect these contributions to appear \cite{Bern:2010tq}.

If we restrict the loop integration measure in \eqn{integranddefinition} to a four-dimensional subspace, as when interpreting the two extra dimensions as a massive regulator, the measure will provide an extra inversion weight of $\prod_i (x_{l_i}^2)^{-4}$, which exactly cancels the extra weight of the integrand. The inversion then commutes with the integration, since the infrared singularities have been regulated, and the amplitude obeys an exact dual conformal symmetry to all loops, which we may write as
\begin{equation}
  I\left[\int \Biggl( \prod_{i=1}^{L}d^4 x_{l_i} \Biggr) \mathcal{I}_n^{L}\right] = 
    \left(\prod_{i\in\E} x_i^2\right) \int \Biggl( \prod_{i=1}^{L}d^4 x_{l_i} \Biggr) \mathcal{I}_n^{L}
   \,, \hskip 1.0cm \hbox{(massively regulated $\NeqFour$)} \,.
\end{equation}
%

\section{Conclusion}

In this paper we demonstrated that the six-dimensional maximal sYM tree-level amplitudes and multiloop integrands exhibit dual conformal covariance. While dual conformal symmetry has been shown to exist for theories in $D\neq4$~\cite{Huang:2010qy}, it is noteworthy that such a symmetry can be defined for theories which are not invariant under ordinary conformal symmetry~\cite{Bern:2010qa,CaronHuot:2010rj}. Also, because a massless on-shell particle in six dimensions is equivalent to a massive particle in four dimensions, our six-dimensional result then naturally gives the dual conformal properties of the massively regulated four-dimensional $\NeqFour$ theory~\cite{Alday:2009zm}.

Covariance under dual conformal symmetry facilitated the construction of four-dimensional $\NeqFour$ sYM tree-level amplitudes by expressing the amplitudes in terms of dual conformal invariant ``$R$" functions~\cite{Drummond:2008cr}. Therefore, an obvious task is to formulate the corresponding ``$R$" covariants for the six-dimensional theory and construct the general $n$-point tree amplitude. This would serve as an efficient way to compute massive amplitudes in four dimensions.

One of the important new ingredients in utilizing the four-dimensional dual conformal symmetry to determine amplitudes is the notion of momentum twistors~\cite{Hodges:2009hk}. These are twistor variables whose incidence relations are defined in the dual momentum space instead of the ordinary spacetime. Similarly, one can now hope to express six-dimensional amplitudes in terms of momentum twistors defined in six dimensions. It would be interesting to see if such a construction leads to alternative representations of planar maximal sYM amplitudes.

\section{Acknowledgements}
We thank Zvi Bern for suggesting this problem and for many
stimulating discussions. We would also like to thank Donal O'Connell and  Emery Sokatchev for their useful communications. This research was supported by the US
Department of Energy under contract DE--FG03--91ER40662. T.D. gratefully acknowledges the financial support of a Robert Finkelstein grant.

\appendix
\section{Clifford algebra conventions}
Here we follow the conventions of~\cite{CheungOConnell}. The Clifford algebra is given as,
\begin{equation}
\sigma^\mu_{AB}\tilde{\sigma}^{\nu BC}+\sigma^\nu_{AB}\tilde{\sigma}^{\mu BC}=2\eta^{\mu\nu}\delta^C_A\,.
\end{equation}
The explicit forms of the matrices $\sigma, \tilde{\sigma}$ are given in ~\cite{CheungOConnell}. They satisfy the following identities:
\begin{eqnarray}
\nonumber&&\sigma^\mu_{AB}\sigma_{\mu CD}=-2\epsilon_{ABCD} \,,\\
\nonumber&&\tilde{\sigma}^{\mu AB}\tilde{\sigma}_{\mu}^{CD}=-2\epsilon^{ABCD} \,,\\
\nonumber&&\tilde{\sigma}^{\mu AB}\sigma_{\mu CD}=-2\left(\delta_C^{[A}\delta_{D}^{B]}\right) \,,\\
&&tr(\sigma^\mu\tilde{\sigma}_\nu)=\sigma^{\mu}_{AB}\tilde{\sigma}^{\nu BA}=4\eta^{\mu\nu}\,.
\end{eqnarray}
From the above, one can deduce, 
\begin{eqnarray}
&&x^\mu=\frac{1}{4}(\tilde{\sigma}^\mu)^{BA}x_{AB} = \frac{1}{4}(\sigma^\mu)_{BA}x^{AB} \,,\nn\\
&&x^{AB}=\frac{1}{2}\epsilon^{ABCD}x_{CD}\,, \nn\\
&&x^{AB}x_{BE}=\frac{1}{2}\epsilon^{ABCD}x_{CD}x_{BE}=x^2\delta^A_E\,,\nn\\
&&x^2=x^\nu x_\nu=-\frac{1}{8}\epsilon_{ABCD}x^{AB}x^{CD}\,.
\end{eqnarray}
Some useful formul\ae{}:
\begin{eqnarray}
\nonumber X_{[ab]}&=&\epsilon_{ab}X^c\,_c,\;X^{[ab]}=-\epsilon^{ab}X^c\,_c\\
\nonumber\epsilon^{ABCD}\epsilon_{AEFG}&=&\left(\delta^B_E\delta^C_F\delta^D_G+\delta^C_E\delta^D_F\delta^B_G+\delta^D_E\delta^B_F\delta^C_G\right.\label{econtraction}\\
&&\left.-\delta^B_F\delta^C_E\delta^D_G-\delta^C_F\delta^D_E\delta^B_G-\delta^D_F\delta^B_E\delta^C_G\right)
\end{eqnarray}
%

\section{Proof of variable inversion formul\ae{}}
\label{InversionAppendix}

In this appendix, we derive the inversion properties in \eqn{moreinversions}. 

\begin{itemize}
\item $I[\lambda_{\{ij\}a}^A] = \frac{x_{iAB}\lambda_{\{ij\}}^{Ba}}{\sqrt{x_i^2x_j^2}} = \frac{x_{jAB}\lambda_{\{ij\}}^{Ba}}{\sqrt{x_i^2x_j^2}}$
     
Our starting point is the constraint equation $(x_i-x_j)_{AB} = \tilde\lambda_{\{ij\}A\da}\tilde\lambda_{\{ij\}B}^{\da}\,$. Contracting both sides with $\lambda_{\{ij\}}^{Ba}$ provides an equation linear in $\lambda$, but loses normalization information. We then proceed with the inversion
\begin{eqnarray}
  0 &=& I[(x_i-x_j)_{AB}\lambda_{\{ij\}}^{Ba}] \nn\\
  &=& (x_j^{-1})^{AC}(x_i-x_j)_{CD}(x_i^{-1})^{DB} I[\lambda_{\{ij\}}^{Ba}].
\end{eqnarray}
This implies that $(x_{i}^{-1})^{DB} I[\lambda_{\{ij\}}^{Ba}]$ is in the null space of $(x_i-x_j)_{CD}$, so 
\begin{eqnarray}
  (x_{i}^{-1})^{DB} I[\lambda_{\{ij\}}^{Ba}] &=& M_{ab}\lambda_{\{ij\}}^{Db} \nn\\
  \Rightarrow \quad I[\lambda_{\{ij\}}^{Aa}] &=& x_{iAB} M_{ab} \lambda_{\{ij\}}^{Bb} \nn\\
  &=& x_{jAB}M_{ab}\lambda_{\{ij\}}^{Bb}\,,
\end{eqnarray}
where $M_{ab}$ is a normalization matrix, which we partially fix by inverting the original constraint equation
\begin{eqnarray}
  &&I[(x_i-x_j)^{AB}] = I[\lambda_{\{ij\}}^{Aa}] I[\lambda_{\{ij\}a}^B] \nn\\
 \Rightarrow&& (x_i^{-1})_{AC}(x_i-x_j)^{CD}(x_j^{-1})_{DB} = x_{iAC} M_{ab}\lambda_{\{ij\}}^{Cb} \, x_{jBD}M^{ac}\lambda_{\{ij\}c}^D \nn\\
 \Rightarrow&& (x_i-x_j)^{CD} = -x_i^2 x_j^2 M_{ab}M^{ac} \lambda_{\{ij\}}^{Cb} \lambda_{\{ij\}c}^{D} \nn\\
 \Rightarrow&& M_{ab}M^{ac} = -\frac{\delta_b^c}{x_i^2x_j^2}\,.
\end{eqnarray}
This is the only constraint on $M$. Without loss of generality, we choose $M_{ab}=\epsilon_{ab}(x_i^2x_j^2)^{-1/2}$.

\item $I[\eta_{\{ij\}a}] = -\sqrt{\frac{x^2_i}{x^2_j}} \Bigl(\eta^a_{\{ij\}}+(x^{-1}_i)_{AB}\theta^A_i\lambda_{\{ij\}}^{Ba}\Bigr)$

To invert $\eta$, we begin with the constraint equation $(\theta_i-\theta_j)^A = \lambda_{\{ij\}}^{Aa}\eta_{\{ij\}a}\,$. Inverting, we have
\begin{eqnarray}
  &&I[(\theta_i-\theta_j)^A] = I[\lambda_{\{ij\}}^{Aa}]I[\eta_{\{ij\}a}] \nn\\
  \Rightarrow&& (x_i^{-1})_{AB}\,\theta_i^B - (x_j^{-1})_{AB}\,\theta_j^B = 
    \frac{x_{iAB}}{\sqrt{x_i^2x_j^2}}\lambda_{\{ij\}a}^{B}I[\eta_{\{ij\}a}] \nn\\
  && \phantom{ (x_i^{-1})_{AB}\,\theta_i^B - (x_j^{-1})_{AB}\,\theta_j^B} = 
    \frac{x_{jAB}}{\sqrt{x_i^2x_j^2}}\lambda_{\{ij\}a}^{B}I[\eta_{\{ij\}a}]\,.
\end{eqnarray}
Multiplying the above equations by $x_i$ and $x_j$, respectively, we get
\begin{eqnarray}
  \theta_i^A - (x_ix_j^{-1})^A_{\phantom{A}B}\,\theta_j^B &=& \sqrt{\frac{x_i^2}{x_j^2}} \lambda_{\{ij\}a}^{A}I[\eta_{\{ij\}a}]\,, \nn\\
  (x_jx_i^{-1})^A_{\phantom{A}B}\,\theta_i^B - \theta_j^A &=& \sqrt{\frac{x_j^2}{x_i^2}} \lambda_{\{ij\}a}^{A}I[\eta_{\{ij\}a}]\,.
\end{eqnarray}
Adding these two equations gives
\begin{eqnarray}
  (\theta_i-\theta_j)^A - (x_ix_j^{-1})^A_{\phantom{A}B}\,\theta_j^B + (x_jx_i^{-1})^A_{\phantom{A}B}\,\theta_i^B 
    = \frac{x_i^2+x_j^2}{\sqrt{x_i^2x_j^2}}\lambda_{\{ij\}a}^{A}I[\eta_{\{ij\}a}]\,.
\end{eqnarray}
We rewrite the LHS as
\begin{eqnarray}
&&\hskip -2.0cm  -\lambda_{\{ij\}a}^A\eta_{\{ij\}}^{a} - \frac{(x_ix_j)^A_{\phantom{A}B}\,\theta_j^B}{x_j^2}
    +\frac{(x_jx_i)^A_{\phantom{A}B}\,\theta_{i}^B}{x_i^2} \nn\\
&& =-\lambda_{\{ij\}a}^A\eta_{\{ij\}}^{a}
  - \frac{(x_ix_j)^A_{\phantom{A}B}\,\theta_{i}^B - x_i^2(\theta_i-\theta_j)^A}{x_j^2}
  + \frac{(x_jx_i)^A_{\phantom{A}B}\,\theta_{i}^B}{x_i^2} \nn\\
&& = -\frac{x_i^2+x_j^2}{x_j^2}\left(\lambda_{\{ij\}a}^A\eta_{\{ij\}}^{a} 
    + \theta_i^A - (x_jx_i^{-1})^A_{\phantom{A}B}\,\theta_{i}^B \right) \nn\\
&& = -\frac{x_i^2+x_j^2}{x_j^2}\left(\lambda_{\{ij\}a}^A\eta_{\{ij\}}^{a} 
    +(x_i-x_j)^{AB}(x_i^{-1})_{BC}\,\theta_i^C\right) \nn\\
&& = -\frac{x_i^2+x_j^2}{x_j^2}\lambda_{\{ij\}a}^A\left(\eta_{\{ij\}}^{a} 
    -\lambda^{Ba}_{\{ij\}}(x_i^{-1})_{BC}\,\theta_i^C\right) \,.
\end{eqnarray}
where in the second line we used $(x_i-x_j)_{AB}(\theta_i-\theta_j)^B=0\,$, and in the third line we used $(x_i-x_j)^{AB}(x_i-x_j)_{BC} = 0\,$. We can now read off the solution.

\item $I[u_{ia}]=\frac{\beta u_i^a}{\sqrt{x^2_{i+2}}}\,, \quad I[\tilde{u}_{i\da}]=\frac{\tilde{u}_i^{\da}}{\beta\sqrt{x^2_{i+2}}}$

Here, we begin with the definition $\langle i_a |i+1_{\db}] = u_{ia}\tilde{u}_{i+1\db}$. Contracting both sides with $u_i^a$ and inverting, we have
\begin{eqnarray}
  I[u_i^a] I\left[\langle i_a|i+1_{\db}]\right] = -I[u_i^a] \frac{\langle i^a|i+1^{\db}]}{\sqrt{x_i^2x_{i+2}^2}} = 0\,.
\end{eqnarray}
Since $u_{ia}$ is the only vector annihilated by the matrix $\langle i|i+1]$, we conclude that
\begin{equation}
  I[u_i^a] = \alpha_i u_{ia}\,,
\end{equation}
for some $\alpha_i$. Returning to the original equation defining $u$ and $\tilde{u}$, we get a set of constraints on $\alpha_i$,
\begin{align}
  \alpha_1 \tilde\alpha_2 &= -(x_1^2 x_3^2)^{-1/2}\,, & \tilde\alpha_1\alpha_2 =& -(x_1^2 x_3^2)^{-1/2}\,, \nn\\
  \alpha_2 \tilde\alpha_3 &= -(x_2^2 x_1^2)^{-1/2}\,, & \tilde\alpha_2\alpha_3 =& -(x_2^2 x_1^2)^{-1/2}\,, \nn\\
  \alpha_3 \tilde\alpha_1 &= -(x_3^2 x_2^2)^{-1/2}\,, & \tilde\alpha_3\alpha_1 =& -(x_3^2 x_2^2)^{-1/2} \,.
\end{align}
The solution to these equations is
\begin{align}
  \alpha_i &= \frac{\beta}{\sqrt{x_{i+2}^2}}\,, & \tilde\alpha_i = \frac{1}{\beta\sqrt{x_{i+2}^2}}\,.
\end{align}

\item $I[w_{ia}]=-\frac{1}{\beta}\sqrt{x^2_{i+2}}w_i^{a}\,, \quad
  I[\tilde{w}_{i\dot{a}}]=-\beta\sqrt{x^2_{i+2}}\tilde{w}_i^{\dot{a}}$

Because $w$ is defined as the pseudoinverse of $u$ via $u_{ia}w_{ib}-u_{ib}w_{ia}=\epsilon_{ab}$, its inversion is straightforward. The definition inverts as
\begin{eqnarray}
  \epsilon^{ba} &=& I[u_{ia}]I[w_{ib}]-I[u_{ib}]I[w_{ia}] \,, \nn\\
  &=& \frac{\beta}{\sqrt{x_{i+2}^2}} \left(u_i^aI[w_{ib}]-u_i^bI[w_{ia}]\right) \,.
\end{eqnarray}
This is again the definition of $w$ as the psuedoinverse of $u$,
\begin{equation}
  \frac{\beta}{\sqrt{x_{i+2}^2}}I[w_{ia}] = -w_i^a \,,
\end{equation}
whence the result follows.

\end{itemize}



\begin{thebibliography}{99}
\bibitem{DualConformal}
J.~M.~Drummond, J.~Henn, V.~A.~Smirnov and E.~Sokatchev,
JHEP {\bf 0701}, 064 (2007)
[hep-th/0607160];\\
%
\bibitem{DualConformal2}
J.~M.~Drummond, J.~Henn, G.~P.~Korchemsky and E.~Sokatchev,
Nucl.\ Phys.\  B {\bf 828}, 317 (2010)
[0807.1095 [hep-th]].
%

\bibitem{Strong}
L.~F.~Alday and J.~Maldacena,
JHEP {\bf 0706}, 064 (2007)
[0705.0303 [hep-th]],\\
%
 R.~Ricci, A.~A.~Tseytlin and M.~Wolf,
  JHEP {\bf 0712}, 082 (2007)
  [arXiv:0711.0707 [hep-th]],\\
N.~Beisert, R.~Ricci, A.~A.~Tseytlin and M.~Wolf,
  Phys.\ Rev.\  D {\bf 78}, 126004 (2008)
  [arXiv:0807.3228 [hep-th]],\\
N.~Berkovits and J.~Maldacena,
  JHEP {\bf 0809}, 062 (2008)
  [arXiv:0807.3196 [hep-th]]
\bibitem{Weak}
G.~P.~Korchemsky, J.~M.~Drummond and E.~Sokatchev,
  Nucl.\ Phys.\  B {\bf 795}, 385 (2008)
  [arXiv:0707.0243 [hep-th]];\\
  A.~Brandhuber, P.~Heslop and G.~Travaglini,
  Nucl.\ Phys.\  B {\bf 794}, 231 (2008)
  [arXiv:0707.1153 [hep-th]].\\
\bibitem{anomaly}
J.~M.~Drummond, J.~Henn, G.~P.~Korchemsky and E.~Sokatchev,
Nucl.\ Phys.\  B {\bf 795}, 52 (2008)
[0709.2368 [hep-th]];\\
J.~M.~Drummond, J.~Henn, G.~P.~Korchemsky and E.~Sokatchev,
Nucl.\ Phys.\  B {\bf 826}, 337 (2010)
[0712.1223 [hep-th]].
\bibitem{Drummond:2009fd}
  J.~M.~Drummond, J.~M.~Henn and J.~Plefka,
  JHEP {\bf 0905}, 046 (2009)
  [arXiv:0902.2987 [hep-th]].
\bibitem{Drummond:2008cr}
 J.~M.~Drummond and J.~M.~Henn,
  JHEP {\bf 0904}, 018 (2009)
  [arXiv:0808.2475 [hep-th]];
\bibitem{Bern:2006ew}
Z.~Bern, M.~Czakon, L.~J.~Dixon, D.~A.~Kosower and V.~A.~Smirnov,
  Phys.\ Rev.\  D {\bf 75}, 085010 (2007)
  [arXiv:hep-th/0610248];\\
Z.~Bern, J.~J.~M.~Carrasco, H.~Johansson and D.~A.~Kosower,
  Phys.\ Rev.\  D {\bf 76}, 125020 (2007)
  [arXiv:0705.1864 [hep-th]];\\
  Z.~Bern, L.~J.~Dixon, D.~A.~Kosower, R.~Roiban, M.~Spradlin, C.~Vergu and A.~Volovich,
  Phys.\ Rev.\  D {\bf 78}, 045007 (2008)
  [arXiv:0803.1465 [hep-th]];\\
\bibitem{Korchemsky:2010ut}
  G.~P.~Korchemsky and E.~Sokatchev,
  Nucl.\ Phys.\  B {\bf 839}, 377 (2010)
  [arXiv:1002.4625 [hep-th]].
\bibitem{ArkaniHamed:2010kv}
  N.~Arkani-Hamed, J.~L.~Bourjaily, F.~Cachazo, S.~Caron-Huot and J.~Trnka,
  arXiv:1008.2958 [hep-th];\\
  J.~M.~Drummond and J.~M.~Henn,
  arXiv:1008.2965 [hep-th].

\bibitem{Hodges:2009hk}
  A.~Hodges,
  arXiv:0905.1473 [hep-th].
\bibitem{ArkaniHamed:2009vw}
  N.~Arkani-Hamed, F.~Cachazo and C.~Cheung,
  JHEP {\bf 1003}, 036 (2010)
  [arXiv:0909.0483 [hep-th]].
\bibitem{Mason:2009qx}
  L.~Mason and D.~Skinner,
  JHEP {\bf 0911}, 045 (2009)
  [arXiv:0909.0250 [hep-th]].

\bibitem{anomaly2}
  A.~Brandhuber, P.~Heslop and G.~Travaglini,
  JHEP {\bf 0908}, 095 (2009)
  [arXiv:0905.4377 [hep-th]];JHEP {\bf 0910}, 063 (2009)
  [arXiv:0906.3552 [hep-th]];\\
N.~Beisert, J.~Henn, T.~McLoughlin and J.~Plefka,
  JHEP {\bf 1004}, 085 (2010)
  [arXiv:1002.1733 [hep-th]].





\bibitem{Alday:2009zm}
  L.~F.~Alday, J.~M.~Henn, J.~Plefka and T.~Schuster,
  JHEP {\bf 1001}, 077 (2010)
  [arXiv:0908.0684 [hep-th]];\\
  J.~M.~Henn, S.~G.~Naculich, H.~J.~Schnitzer and M.~Spradlin,
  JHEP {\bf 1004}, 038 (2010)
  [arXiv:1001.1358 [hep-th]];\\
 J.~M.~Henn,
  Nucl.\ Phys.\ Proc.\ Suppl.\  {\bf 205-206}, 193 (2010)
  [arXiv:1005.2902 [hep-ph]].
\bibitem{Bern:2010qa}
  Z.~Bern, J.~J.~Carrasco, T.~Dennen, Y.~t.~Huang and H.~Ita,
  arXiv:1010.0494 [hep-th].


\bibitem{CheungOConnell}
C.~Cheung and D.~O'Connell,
JHEP {\bf 0907}, 075 (2009)
[0902.0981 [hep-th]];\\
 \bibitem{DHS}
T.~Dennen, Y.~t.~Huang and W.~Siegel,
JHEP {\bf 1004}, 127 (2010)
[0910.2688 [hep-th]].

 
\bibitem{BCFW}
 R.~Britto, F.~Cachazo and B.~Feng,
  Nucl.\ Phys.\  B {\bf 715}, 499 (2005)
  [arXiv:hep-th/0412308];\\
R.~Britto, F.~Cachazo, B.~Feng and E.~Witten,
  Phys.\ Rev.\ Lett.\  {\bf 94}, 181602 (2005)
  [arXiv:hep-th/0501052].
\bibitem{Brandhuber:2008pf}
  A.~Brandhuber, P.~Heslop and G.~Travaglini,
  Phys.\ Rev.\  D {\bf 78}, 125005 (2008)
  [arXiv:0807.4097 [hep-th]].

\bibitem{CaronHuot:2010rj}
  S.~Caron-Huot and D.~O'Connell,
  arXiv:1010.5487 [hep-th].

\bibitem{UnitarityMethod}
Z.~Bern, L.~J.~Dixon, D.~C.~Dunbar and D.~A.~Kosower,
Nucl.\ Phys.\  B {\bf 425}, 217 (1994)
[hep-ph/9403226];
%
Nucl.\ Phys.\  B {\bf 435}, 59 (1995)
[hep-ph/9409265];\\
%
Z.~Bern, L.~J.~Dixon and D.~A.~Kosower,
Ann.\ Rev.\ Nucl.\ Part.\ Sci.\  {\bf 46}, 109 (1996)
[hep-ph/9602280]; JHEP {\bf 0408}, 012 (2004)
  [arXiv:hep-ph/0404293].
  

  
\bibitem{Boels:2009bv}
  R.~Boels,
  JHEP {\bf 1001}, 010 (2010)
  [arXiv:0908.0738 [hep-th]].

\bibitem{Brandhuber:2010mm}
  A.~Brandhuber, D.~Korres, D.~Koschade and G.~Travaglini,
  arXiv:1010.1515 [hep-th].
  


\bibitem{Huang:2010rn}
T.~Chern,
  arXiv:0906.0657 [hep-th];\\
  Y.~t.~Huang and A.~Lipstein,
  JHEP {\bf 1010}, 007 (2010)
  [arXiv:1004.4735 [hep-th]].


\bibitem{Bern:2010tq}
  Z.~Bern, J.~J.~M.~Carrasco, L.~J.~Dixon, H.~Johansson and R.~Roiban,
  arXiv:1008.3327 [hep-th].


\bibitem{Huang:2010qy}
T.~Bargheer, F.~Loebbert and C.~Meneghelli,
  Phys.\ Rev.\  D {\bf 82}, 045016 (2010)
  [arXiv:1003.6120 [hep-th]].
 S.~Lee,
  Phys.\ Rev.\ Lett.\  {\bf 105}, 151603 (2010)
  [arXiv:1007.4772 [hep-th]];\\
  Y.~t.~Huang and A.~E.~Lipstein,
  JHEP {\bf 1011}, 076 (2010)
  [arXiv:1008.0041 [hep-th]].


\end{thebibliography}
\end{document}